# High spin polarization in the ferromagnetic filled skutterudites $KFe_4Sb_{12}$ and $NaFe_4Sb_{12}$

Goutam Sheet,[1] H. Rosner,[2] S. Wirth,[2] A. Leithe-Jasper,[2] W. Schnelle,[2]
U. Burkhardt,[2] J. A. Mydosh,[3,2] P. Raychaudhuri,[1] and Yu. Grin[2]

[1]*Department of Condensed Matter Physics and Materials Science, Tata Institute of Fundamental Research, Homi Bhabha Road, Colaba, Mumbai 400005, India*
[2]*Max Planck Institute for Chemical Physics of Solids, Nöthnitzer Straße 40, 01187 Dresden, Germany*
[3]*Kamerlingh Onnes Laboratory, Leiden University, Leiden, The Netherlands*

The spin polarization of ferromagnetic alkali-metal iron antimonides $KFe_4Sb_{12}$ and $NaFe_4Sb_{12}$ is studied by point-contact Andreev reflection using superconducting Nb and Pb tips. From these measurements an intrinsic transport spin polarization $P_t$ of 67% and 60% for the K and Na compound, respectively, is inferred which establishes these materials as a new class of highly spin polarized ferromagnets. The results are in accord with band structure calculations within the local spin density approximation (LSDA) that predict nearly 100% spin polarization in the density of states. We discuss the impact of calculated Fermi velocities and spin fluctuations on $P_t$.

Electronic devices based upon spin control (spintronics)[1] and new materials with a high degree of spin polarization, e.g., half-metallic ferromagnets,[2,3] have evolved as topics of particular attention in recent years. Independently, the filled skutterudite compounds, $AM_4X_{12}$ ($M$ is a transition metal, $X$ a pnictide), have become materials of vast possibilities.[4] For, these compounds can be synthesized with a variety of filler elements, $A$, ranging from sodium spanning the alkaline and rare earths, to the actinides. Usually, the transition metal site (Fe, Ru or Os) is non-magnetic. The lattice parameter of these cubic compounds can be increased systematically by proceeding from P to Sb and from Fe to Os, thereby generating such diverse behavior as unconventional superconductivity, quantum and correlated electron magnetism and large thermoelectric effects,[4–7] the latter initiating the interest in these materials.

Particular interesting examples are the filled skutterudites with iron-antimony host.[8] Here, the magnetic properties are governed by the charge of the filler ion, especially if it is non-magnetic, e.g., $Na^{1+}$, $Ca^{2+}$, $La^{3+}$. Surprisingly, the compounds containing monovalent cations are ferromagnetic. In our prior work[9,10] we have evidenced that the Na and K compounds are weak itinerant ferromagnets with a Curie temperature $T_C = 85\,K$ where a clear magnetic phase transition is observed in all bulk properties. Previous band-structure calculations within the local spin density approximation (LSDA) for $NaFe_4Sb_{12}$ predicted a spin-split density of states with almost perfect half-metallic behavior.[10] Compounds with divalent fillers (e.g. $CaFe_4Sb_{12}$) are nearly ferromagnetic metals, i.e. they do not show ferromagnetism. Yet, they exhibit large paramagnetic susceptibilities or Stoner factors and Sommerfeld-Wilson ratios $R_W \approx 24$.[10,12,13]

In this report, we present the first point-contact Andreev reflection (PCAR) measurements[14,15] conducted on $KFe_4Sb_{12}$ and $NaFe_4Sb_{12}$ samples to verify the theoretical prediction. In addition, we present detailed band structure calculations, including the Fermi surface and Fermi velocities. From the combination of PCAR measurements and LSDA calculation we not only infer a remarkably high degree of transport spin polarization $P_t \approx 67\%$, which is a key parameter in spintronics applications.[16] But we can also unambiguously relate this to a high density of states in the metallic channel which protects the spin polarization in these materials against detrimental influences of scattering.[17] This certainly makes $KFe_4Sb_{12}$ and $NaFe_4Sb_{12}$ interesting candidates for spectroscopic investigations. Obviously, these two compounds introduce a new class of materials that exhibits a large degree of spin polarization and itinerant electron magnetism with a rather high $T_C$.

Polycrystalline samples of $KFe_4Sb_{12}$ and $NaFe_4Sb_{12}$ along with $CaFe_4Sb_{12}$ as a non-ferromagnetic reference material were prepared and characterized as described previously.[9,10] Powders were compacted by spark plasma sintering (SPS) at pressures of 600 MPa in an argon atmosphere. All experiments were conducted on pieces cut from the same batches. A polished surface of the $KFe_4Sb_{12}$ sample shows an average grain size of $\approx 15\,\mu m^2$. The electrical resistivities $\rho(300\,K)$ range between 220 and $530\,\mu\Omega\,cm$. As the achieved density of these sinter materials may slightly vary the residual resistivities are sample dependent. Nonetheless, the values of $\rho$ are similar to those of $AFe_4Sb_{12}$ samples with $A$ = Ca, Sr, Ba prepared by the same method.[12] The common feature in all $\rho(T)$ curves is a shoulder around 70–80 K which is a fingerprint of carriers scattering on spin fluctuations.[12,18] On top of this a tiny kink at $\approx 85\,K$ signals the ferromagnetic ordering of the Na and the K compound (cf. [18] for details). The ferromagnetic state of Na and $KFe_4Sb_{12}$ is characterized by a small remanent magnetization $M_r$/Fe-atom $\approx 0.25\,\mu_B$ at 1.8 K.[9,10] Hysteresis curves are already closed at 10 kOe. However, the magnetization keeps increasing up to the maximum applied field ($M$/Fe-atom $\approx 0.60\,\mu_B$ at 140 kOe) indicat-ing the strong spin fluctuations in these ferromagnets.[10] Interestingly, the K and Na compounds exhibit almost identical properties.[9,10]

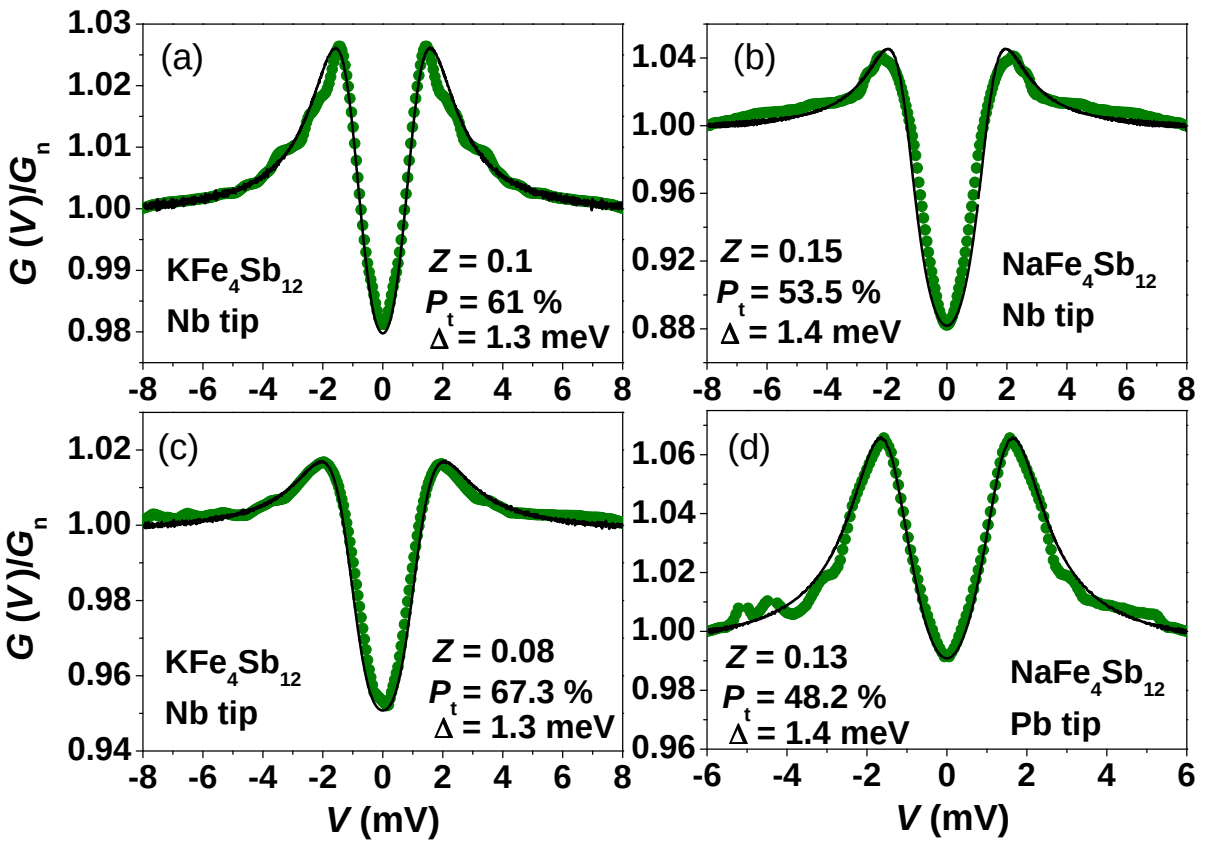

FIG. 1: (color online) Selected conductance spectra for the skutterudites (a), (c) $KFe_4Sb_{12}$ and (b), (d) $NaFe_4Sb_{12}$. The spectra were measured (dots) using (a-c) Nb or (d) Pb tips and normalized to the normal state conductance $G_n$ at high voltage. The lines present the results of the best fits with the fit parameters given in each panel (see text).

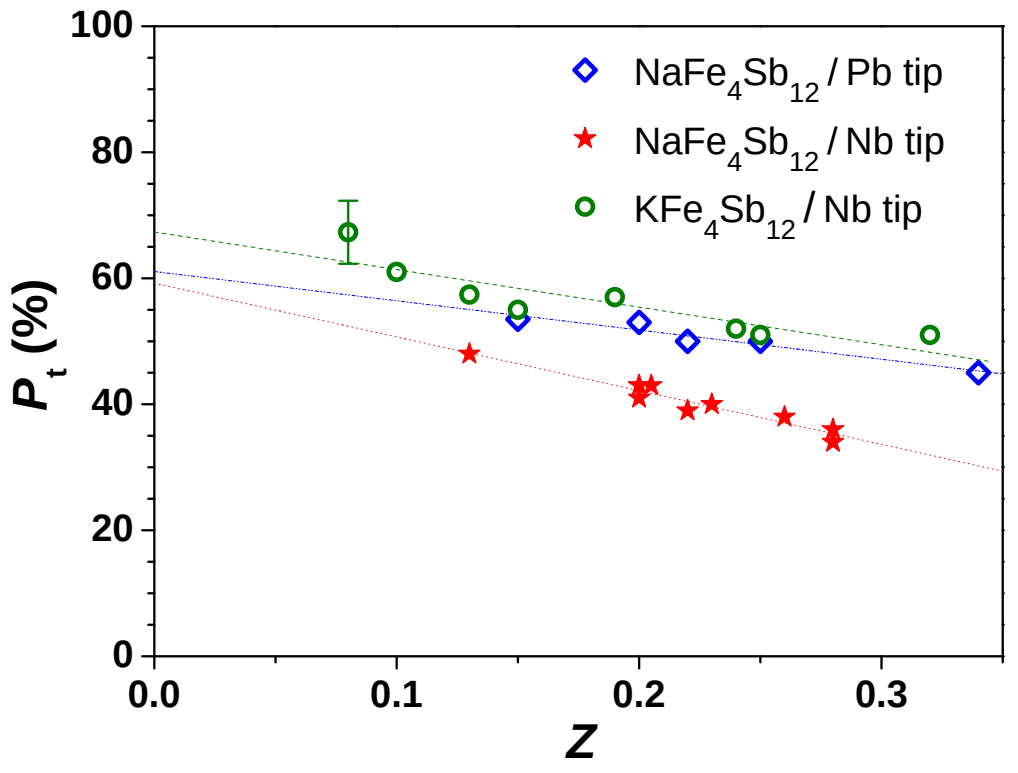

FIG. 2: (color) Transport spin polarization $P_t$ in dependence on the $Z$-value for the different sample/tip combinations measured. The extrapolation to zero $Z$ yields the intrinsic spin polarization of the two skutterudites. The indicated error bar is representative for all measurements.

The point-contact Andreev reflection measurements were performed at 2.8 K in a liquid He continuous flow cryostat. The samples were polished with very fine emery paper to a mirror finish and immediately loaded for experiments to avoid surface degradation. Mechanically cut sharp Nb or Pb tips were engaged on the samples by a differential screw arrangement to establish contacts of minimum size. Differential conductance $G(V)$ versus voltage $V$ characteristics of the contact were obtained using a lock-in modulation technique at 372 Hz. The typical resistance of the contacts in the normal state varied between $10-20\ \Omega$.

In Fig. 1 we show four representative spectra with various combinations of ferromagnetic samples and superconducting tips. The transport spin polarization $P_t$ was extracted from the spectra by fitting a Blonder-Tinkham-Klapwijk (BTK) theory[19] modified to incorporate the effect of spin polarization.[20,21] The vast majority of spectra were analyzed using three fitting parameters, namely $P_t$, the superconducting energy gap $\Delta$, and the barrier parameter $Z$ which characterizes the strength of the potential barrier at the interface. While modeling the spectra, $\Delta$ was kept within 10% of its bulk value for the given superconductors. For a small number of spectra we added to $\Delta$ a small broadening parameter, $\Gamma \sim 0.1-0.2$ meV, for fine corrections to the fits. However, no $\Gamma$ was required for fitting the spectra with low $Z$ values (as, e.g, those presented in Fig. 1) which is representative of the intrinsic spin polarization.[22] Note that the comparatively low $Z$ values of these spectra ensure a good quality of the fits and have the highest impact on the extrapolated *intrinsic* $P_t$ values.

Figure 2 shows the variation of the extracted values of $P_t$ as a function of the barrier parameter $Z$. $P_t$ decreases with increasing $Z$, a behavior that has been observed earlier in a variety of ferromagnets and is believed to arise from spin depolarization at a magnetically disordered scattering barrier formed at the interface. The intrinsic value of $P_t$ is therefore extracted by linearly extrapolating the $P_t$ vs. $Z$ curve to $Z=0$. The consistency of this procedure is easily seen in the case of $NaFe_4Sb_{12}$: measurements carried out with both Pb and Nb tips result in the same value of the intrinsic spin polarization (within experimental errors), though the decay of $P_t$ with $Z$ is different for the two cases due to differences in the nature of the interfaces. The intrinsic value of $P_t$ extracted in this way is 67% for $KFe_4Sb_{12}$, and 60% for $NaFe_4Sb_{12}$.[23] We note that PCAR measurements were also conducted on $CaFe_4Sb_{12}$, the structurally closest but non-ferromagnetic homologue to the compounds of interest here. As expected we found $P_t=0$ for all fitting attempts using $CaFe_4Sb_{12}$. This result constitutes a consistency check with respect to the $P_t$ values as extracted from the PCAR measurements.

To gain further insight into the electronic structure of these materials on a microscopic level, a full-potential non-orthogonal local-orbital calculation scheme (FPLO)[24] within the LSDA was utilized. In the scalar-relativistic calculations the exchange and correlation potential of Perdew and Wang[25] was used. As the basis set, Na(2$s$, 2$p$, 3$s$, 3$p$, 3$d$), K(3$s$, 3$p$, 4$s$, 4$p$, 3$d$), Fe (3$s$, 3$p$, 4$s$, 4$p$, 3$d$) and Sb (4$s$, 4$p$, 4$d$, 5$s$, 5$p$, 5$d$) states were employed. The lower-lying states were treated fully relativistically as core states. The Na and K 3$d$ states as well as the Sb 5$d$ states were taken into account as polarization states to increase the completeness of the basis set. The treatment of the Na(2$s$, 2$p$), K(3$s$, 3$p$), Fe(3$s$, 3$p$) and Sb (4$s$, 4$p$, 4$d$) semi-core like states as valence states was necessary to account for non-negligible core-core overlaps. The spatial extension of the basis orbitals, controlled by a confining potential $(r/r_0)^4$, was optimized to minimize the total energy.[26] For self consistency, a fine $k$-mesh of 1256 points in the irreducible part of the Brillouin zone (27,000 in the full zone) was used. To ensure accurate density of states (DOS) and band structure information, especially to obtain smooth Fermi surfaces and reliable Fermi velocities $v_F$, a $k$-mesh of 1,000,000 points was used in a final step.

Our band structure calculations result in a ferromagnetic ground state for $KFe_4Sb_{12}$ with a nearly integer magnetic moment of 2.98 $\mu_B$ per formula unit. The states in the spin split region originate predominantly from Fe-3$d$ hybridized with the Sb-5$p$ states (see Fig. 3). We find an almost fully polarized DOS at the Fermi level $E_F$ for $KFe_4Sb_{12}$ [see Fig. 3(a)]. The DOS contribution in the spin-up channel can be assigned mainly to two bands (Fermi surfaces) of almost pure Fe-3$d$ character [see Fig. 3(b)]. There is only one band of strongly mixed Sb 5$p$–Fe 3$d$ character crossing $E_F$ in the spin-down channel [see Fig. 3(b)]. Due to its high value of $v_F \sim 0.3 \times 10^6$ m s$^{-1}$, this band makes only a tiny contribution to the number of states at $E_F$.

In case of spin polarization ($P_n$) measurements it is imperative to bear in mind that in addition to the DOS at $E_F$, $N(E_F)$, also the Fermi velocity $v_F$ may influence the obtained value of $P_n$. $P_n$ may be defined by[27,28]

$$P_n = \frac{\int N_\uparrow(E_F)\, v_{F\uparrow}^n\, dS - \int N_\downarrow(E_F)\, v_{F\downarrow}^n\, dS}{\int N_\uparrow(E_F)\, v_{F\uparrow}^n\, dS + \int N_\downarrow(E_F)\, v_{F\downarrow}^n\, dS} \qquad (1)$$

where the integrals are taken over all Fermi surfaces and the arrows distinguish the two spin channels. The exponent $n$ is determined by the applied experimental technique (at least in the case of $P_n \neq 1$): Only in the case of spin-resolved photoemission, Eq. (1) simplifies to a form $P_0 = (N_\uparrow - N_\downarrow)/(N_\uparrow + N_\downarrow)$ solely determined by the DOS at $E_F$. This value can, of course, also be obtained from band structure calculations. For our $KFe_4Sb_{12}$ the spin polarization amounts to $P_0 = 99.6\%$, a value very similar to the one reported for the isovalent Na compound.[10] In the case of electronic transport measurements (as our PCAR measurements) a further distinction has to be made depending on the size of the contact $d$ in comparison to the elastic and inelastic electronic mean free path, $l_e$ and $l_{in}$, respectively. In the ballistic regime $l_e > d$ one finds $n = 1$, whereas $n = 2$ in the diffusive regime $l_e < d < l_{in}$ (in the thermal regime, $l_{in} < d$, is not of interest here since all spectral information is lost). Tunneling experiments can also be described by Eq. (1) and $n = 2$ and correspond to large values of $Z$.

From the band structure calculations we find very different $v_F$ for the two spin channels. As a result, the calculated spin polarization is reduced as the exponent $n$ by which $v_F$ enters into Eq. 1 increases: we calculate $P_0 = 0.996$, $P_1 = 0.968$ and $P_2 = 0.765$. This behavior places emphasis on the fact that the high value of $P_t$ for the skutterudites stems from the almost completely spin-polarized DOS at $E_F$, however reduced by the fact that the spin band with lower (higher) DOS has the higher (lower) Fermi velocity.

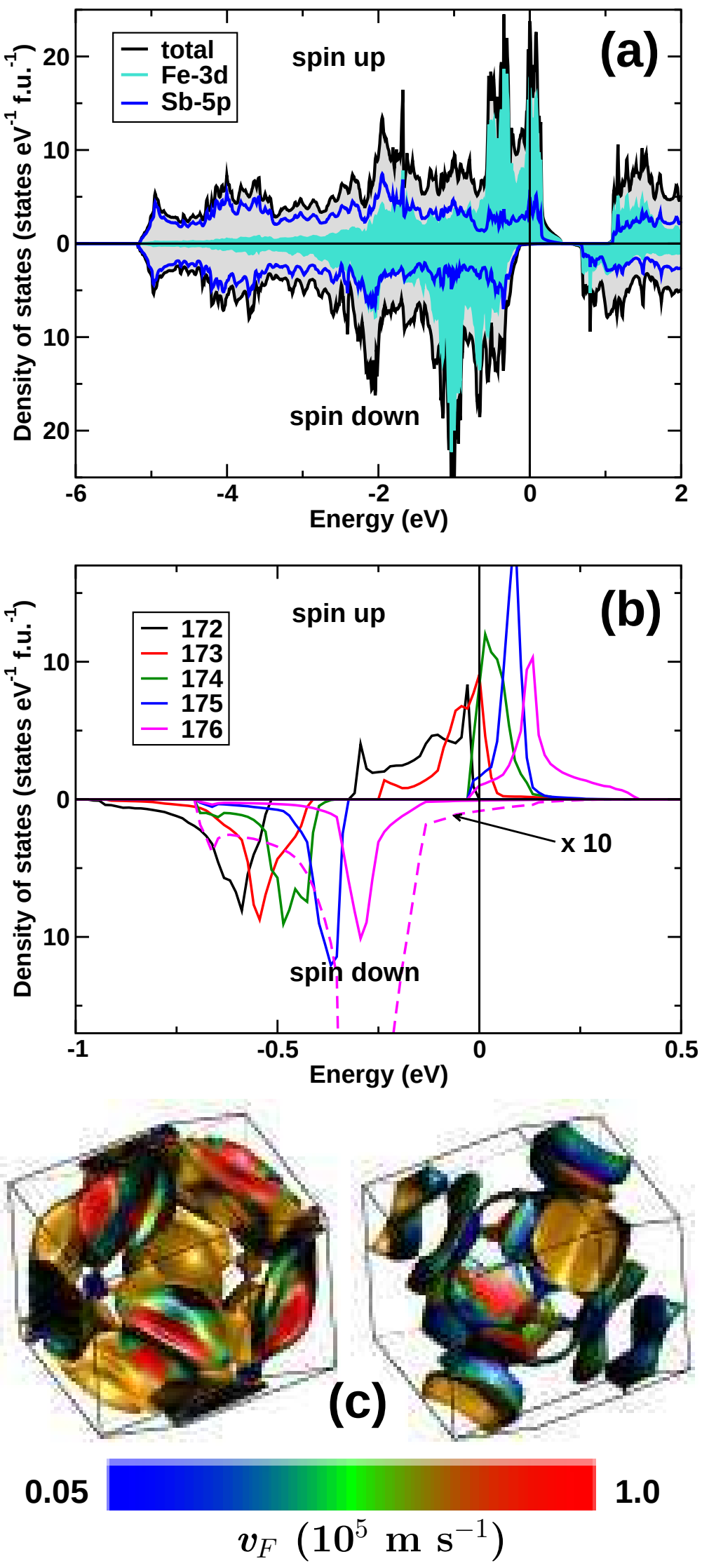

FIG. 3: (color) Calculated density of states (DOS) and Fermi surfaces for $KFe_4Sb_{12}$. (a) Total DOS and partial Fe-3$d$ and Sb-5$p$ contributions. (b) DOS resolved for the individual bands (band numbers are according to the basis set provided in the text) relevant at $E_F$. Spin-down band 176 has a non-vanishing DOS at $E_F$ (cf. dashed line that represents a factor of 10 magnification). (c) Fermi surfaces for the most contributing spin-up bands 173 (left) and 174 (right). The color coding indicates the Fermi velocities $v_F$.

Comparing the experimentally obtained values of the transport spin polarization (67% for $KFe_4Sb_{12}$) with the theoretical calculations in different regimes of transport we observe that though we fitted the spectra using a theory in the ballistic limit, the extracted $P_t$-values are in better agreement with the predictions in the diffusive regime (76.5%). However, it should be noted that the spin polarization as obtained from band structure calculations does not include several factors which are detri-mental to a large transport spin polarization. First, the weak itinerant ferromagnetic nature of the filled skutterudites is associated with large spin fluctuations[9–11] which would reduce the spin polarization from its theoretical value. The presence of strong ferromagnetic fluctuations in these compounds was already indicated by the non-saturating magnetization. Since the high field magnetization value (when spin fluctuations are quenched) is the one corresponding to band structure calculations, a crude correlation between magnetization and spin polarization would suggest a 50% decrease in the spin polarization arising from this effect alone. Second, it has been seen in several compounds[30] that $P_t$ of the surface can be reduced with respect to its bulk value. In addition, spin-orbit coupling in the ferromagnet, although small in our case, reduces the spin polarization at the Fermi level. Third, even if the contacts were in the diffusive regime, this fitting should not yield appreciably different values of the transport spin polarization from the given value assuming a ballistic regime. It has been shown[29] that fitting a spectra from the diffusive regime of point contact by a ballistic model yields approximately the same value of the transport spin polarization (within ∼3%). An estimate of the lower bound of the mean free path within a crystallite of $KFe_4Sb_{12}$, using experimental Hall effect results in a free-electron model, yields 18 nm. Therefore, it is possible that our point contacts with the polycrystalline skutterudite samples were close to the diffusive limit . Considering all these factors the observation of 67% spin polarization in $KFe_4Sb_{12}$ is indeed surprising.

The key to robustness of the transport spin polarization against these detrimental effects may indeed lie in the fact that $P_t$ in this material is primarily governed by the DOS of the majority (spin-down) channel which is very small over a relatively large energy range close to $E_F$. This renders $P_t$ less sensitive not only to fluctuation effects but also with respect to impurity scattering, an important aspect for potential applications.

In conclusion, we have demonstrated by point-contact Andreev reflection measurements large values of transport spin polarization in $KFe_4Sb_{12}$ ($P_t \approx 67\%$) and $NaFe_4Sb_{12}$ ($P_t \approx 60\%$). These results are in line with band structure calculations that link the large $P_t$ to a negligible DOS at $E_F$ in the spin-down channel. The huge DOS at $E_F$ in the minority (spin-up) channel renders these compounds ideal candidates for spectroscopic investigations. Moreover, the filled skutterudites constitute an interesting new class of materials with large potential for spin transport applications.

We are indebted to N. Reinfried, S. Mukhopadhyay, V. Bagwe and S.P. Pai for assistance. G.S. acknowledges TIFR Endowment Fund for partial financial support. H.R. is supported by the Emmy-Noether-Program and J.A.M. by the Alexander-von-Humboldt-Foundation.

[1] S. A. Wolf *et al.*, Science **294**, 1488 (2001).
[2] R. A. de Groot, F. M. Mueller, P. G. van Engen, and K. H. J. Buschow, Phys. Rev. Lett. **50**, 2024 (1983).
[3] W. E. Pickett and J. S. Moodera, Phys. Today **54**(5), 39 (2001).
[4] B. C. Sales, in *Handbook on the Physics and Chemistry of Rare Earths, Vol. 33*, ed. K. A. Gschneidner Jr. (Elsevier, Amsterdam, 2003), 1.
[5] T. Cichorek, A. C. Mota, F. Steglich, N. A. Frederick, W. M. Yuhasz, and M. B. Maple, Phys. Rev. Lett. **94**, 107002 (2005) and references therein.
[6] M. B. Maple, E. D. Bauer, N. A. Frederick, P. C. Ho, W. A. Yuhasz and V. S. Zapf, Physica B **328**, 29 (2003).
[7] B. C. Sales, D. Mandrus, and R. K. Williams, Science **272**, 1325 (1996).
[8] C. Uher, Semiconductors and Semimetals **68**, 139 (2001).
[9] A. Leithe-Jasper *et al.*, Phys. Rev. Lett. **91**, 037208 (2003).
[10] A. Leithe-Jasper *et al.*, Phys. Rev. B **70**, 214418 (2004).
[11] A. A. Gippius *et al.*, J. Magn. Magn. Mater., in print.
[12] E. Matsuoka, K. Hayashi,A. Ikeda,K. Tanaka,T. Takabatake and M. Matsumura, J. Phys. Soc. Jpn. **74**, 1382 (2005).
[13] W. Schnelle *et al.*, Phys. Rev. B **72**, 020402(R) (2005).
[14] R. J. Soulen, Jr. *et al.*, Science **282**, 85 (1998).
[15] S. K. Upadhyay, A. Palanisami, R. N. Louie, and R. A. Buhrman, Phys. Rev. Lett. **81**, 3247 (1998).
[16] J. M. D. Coey and S. Sanvito, J. Phys. D: Appl. Phys. **37**, 988 (2004).
[17] This is different from the case of, e.g., $La_{0.7}Sr_{0.3}Mn_3O$ where the main difference in transport spin polarization $P_t$ stems from a strong disparity of the Fermi velocities of the two spin channels and hence, renders $P_t$ strongly susceptible to disorder, see also Ref. [28].
[18] W. Schnelle *et al.*, to be published.
[19] G. E. Blonder, M. Tinkham, and T. M. Klapwijk, Phys. Rev. B **25**, 4515 (1982).
[20] I. I. Mazin, A. A. Golubov and B. Nadgorny, J. Appl. Phys. **89**, 7576 (2001).
[21] P. Raychaudhuri,A. P. Mackenzie, J. W. Reinier and M. R. Beasley Phys. Rev. B **67**, 020411(R) (2003).
[22] It was shown that incorporating an arbitrary value of $\Gamma$ can result in ambiguities in the extracted value of $P_t$; see, e.g., Y. Bugoslavsky *et al.*, Phys. Rev. B **71**, 104523 (2005).
[23] Alternatively, it has also been suggested that the dependence of $P_t$ and $Z$ is an outcome of wrongly estimating the superconducting energy gap while fitting the PCAR spectra. In that case however, one expects a systematic variation of$P_t$ with the best fit values of $\Delta$, a feature that we do not observe in our experiments. For more details see Ref. [29].
[24] K. Koepernik and H. Eschrig, Phys. Rev. B **59**, 1743 (1999).
[25] J. P. Perdew and Y. Wang, Phys. Rev. B **45**, 13244 (1992).
[26] H. Eschrig, *Optimized LCAO Method and the Electronic Structure of Extended Systems* (Springer, Berlin, 1989).
[27] I. I. Mazin, Phys. Rev. Lett. **83**, 1427 (1999).
[28] B. Nadgorny *et al.*, Phys. Rev. B **63**, 184433 (2001).
[29] G. T. Woods *et al.*, Phys. Rev. B **70**, 054416 (2004).
[30] see, e.g., J. W. Freeland *et al.*, Nature Materials **4**, 62 (2005) and references therein.